\documentclass[11pt,twoside]{article}
\usepackage{asp2004}
\usepackage{psfig}
\usepackage{epsf}
\usepackage{graphicx}
\usepackage{lscape}
\def\spose#1{\hbox to 0pt{#1\hss}}
\def\lta{\mathrel{\spose{\lower 3pt\hbox{$\mathchar"218$}}
     \raise 2.0pt\hbox{$\mathchar"13C$}}}
\def\gta{\mathrel{\spose{\lower 3pt\hbox{$\mathchar"218$}}
     \raise 2.0pt\hbox{$\mathchar"13E$}}}
\def\etal{{\it et al.\ }}

\def\edcomment#1{\iffalse\marginpar{\raggedright\sl#1\/}\else\relax\fi}
\reversemarginpar

\begin{document}
\title{Properties of YMCs Derived From Photometry}

\author{Uta Fritze -- v. Alvensleben}
\affil{Universit\"atssternwarte G\"ottingen, Geismarlandstr. 11, 37083
G\"ottingen, Germany}

\begin{abstract}
I will show that photometry -- if extending over a reasonable
choice of passbands -- can give fairly precise information about 
young star clusters and their evolutionary state. Optical colors
alone are known to leave severe ambiguities due to degeneracies 
between age, metallicity and extinction. High quality photometry 
including $U, B, V$ or $I$, and a NIR band, however, in combination 
with an extensive grid of evolutionary synthesis models for star 
clusters and a dedicated tool to analyse spectral energy 
distributions allows to assess and largely disentangle star 
cluster ages, metallicities, extinction values and, hence, to 
derive their masses. Gaseous emission contributions sensibly 
affect broad band colors during the youngest stages, depending 
on metallicity. Mass functions of young star cluster systems may 
considerably differ in shape from luminosity functions. 
An ESO ASTROVIRTEL project provides multi-color 
photometry for a large number of young, intermediate age 
and old star cluster systems. As a first example I show results 
obtained for NGC 1569.
\end{abstract}
\thispagestyle{plain}

\section{Motivation}

Christy Tremonti has shown in the previous contribution to what precision individual {\bf S}tar {\bf C}luster ({\bf SC}) spectroscopy -- when combined with suitable spectral evolutionary synthesis models -- is able to yield information about star cluster metallicities, ages, extinction values and masses. Clearly high resolution high S/N spectroscopy of individual SCs is the most desirable source of information. In practice, however, this is severely hampered by the faintness and large number of {\bf Y}oung {\bf S}tar {\bf C}lusters ({\bf YSC}s) in distant starburst and interacting galaxies and, of course, even more so for the intrinsically fainter intermediate-age and old SC systems. The strong and fast fading of clusters in early evolutionary stages causes a strong bias towards the youngest and most massive objects in spectroscopic cluster samples. So, in all instances where a reasonable degree of completeness is required -- e.g. to determine SC mass functions, compare cluster populations of different ages, study changes in SC formation efficiencies, cluster destruction processes and timescales -- photometry is the only choice -- already for YSCs in nearby galaxies and more so for all studies of SCs in more distant galaxies, for intrinsically fainter intermediate-age or old SC systems. And it need not be a bad choice, as I'm going to show, provided a reasonable set of passbands is observed with sufficient photometric accuracy.

\section{Modelling Star Clusters}
Our G\"ottingen Evolutionary Synthesis code GALEV is designed to follow the spectral evolution not only of galaxies with {\bf S}tar {\bf F}ormation ({\bf SF}) histories extended in time but also of so-called {\bf S}imple {\bf S}tellar {\bf P}opulations ({\bf SSP}s) like SCs that form all their stars within a short time ($\sim 10^5$ yr) and with the same metallicity. Using complete stellar isochrones from the Padova group including the thermally pulsing AGB phase (TP-AGB) for stars in the mass range 0.9 -- 7 M$_{\odot}$ and stellar model atmosphere spectra from Lejeune {\it et al.} (1997, 1998) we follow the spectral evolution of SSPs with metallicities in the range $-1.7 \leq {\rm [Fe/H]} \leq +0.4$ from the very early stage of 4 Myr all through a Hubble time. In general, we assume a Salpeter IMF from 0.15 -- 70 M$_{\odot}$ with the lowest stellar masses added to the Padova isochrones from calculations by Chabrier \& Baraffe (1997). Folding the synthetic SC spectra with filter functions we obtain the time evolution of SC luminosities and colors from UV through NIR in a large number of filter systems (HST WFPC2, NICMOS, ACS, Johnson, Washington, Stroemgren, ... cf. Schulz {\it et al.} 2002 for details). Luminosities are absolute luminosities for a given initial mass of the model cluster. Model cluster masses decrease with increasing age due to metallicity-dependent stellar evolutionary mass loss. Stellardynamical mass loss dependent on cluster concentration and environment is not included in our models.

Inclusion of the TP-AGB phase is crucial for colors $V-I$ and $V-K$ of SCs: for ages $\geq 10^8$ yr the presence of AGB stars leads to a sudden reddening of $V-I$ by 0.35 mag and of $V-K$ by as much as 1.5 mag at solar metallicity and somewhat less at lower metallicities. Age-dating of YSCs on the basis of their observed $V-I \sim 0.6$ typical for many YSC systems from a comparison with models that do not include the TP-AGB can yield ages too high by factors up to 10. 

For the very youngest SCs that still have hot enough stars to ionise their surrounding gas the gaseous emission gives important contributions to the fluxes in broad band filters with line emission dominating in the optical and continuous emission in the NIR (cf. Kr\"uger {\it et al.} 1995). Our models include both line and continuum emission on the basis of the Lyman continuum photon flux rate calculated for the SSPs using the ${\rm T_{eff} - N_{Lyc}}$ relation for individual stars from Schaerer \& de Koter (1997). This directly gives continuum and hydrogen line fluxes. They are higher at a given age for low metallicity SCs than for higher metallicitiy ones because of the -- on average -- higher ${\rm T_{eff}}$ of the former. Heavy element lines are included with their fluxes relative to ${\rm H_{\beta}}$ taken from Izotov \etal (1994, 1997) and Izotov \& Thuan (1998) for the respective SSP metallicities (see Anders \& FvA 2003 for details).
While for solar metallicity SSPs the gaseous emission contributes already between $\sim 20 - 30~\%$ to the total broad band fluxes in {\sl UBVIJHK} and $\sim 40~\%$ in {\sl R} at ages $\lta 7$ Myr, these gaseous emission contributions are much higher for lower metallicity SSPs. At [Fe/H]$=-1.7$, gaseous flux contributions are as high as $50-65~\%$ in {\sl UBVRJK} and they remain important for ages up to 20 Myr, changing $U-B$ to the red by $\sim 0.2$ mag and $V-I$ to the blue by $\sim -0.8$ mag. For a solar metallicity SSP the color changes due to gaseous emission contributions are small.

\begin{figure}[!h]
\begin{center}
\includegraphics[angle=-90,width=13.cm]{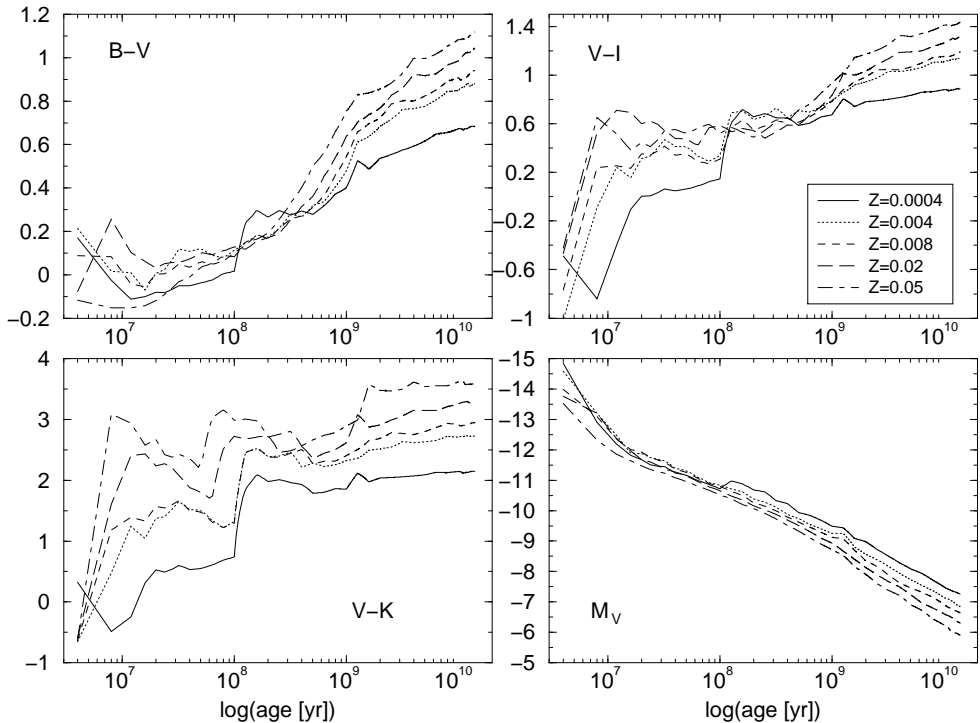}
\end{center}
\caption{Color and luminosity evolution for SSPs of various metallicities.}
\end{figure}

In Fig. 1 we present the color and luminosity evolution for SSPs with 5 different metallicities. Time evolution of both the spectra and photometric properties can be obtained from {\it http://www.uni-sw.gwdg.de/$^{\sim}$galev}. Note that, slightly depending on metallicity, the luminosity evolution, as shown in Fig. 1 for a cluster with the average Milky Way GC mass of ${\rm 3\cdot10^5~M_{\odot}}$, is very strong. A low (high) metallicity cluster fades by 4 (3) mag, respectively, between ages of 4 and 100 Myr. As starburst durations, and hence age differences among YSCs, in massive interacting galaxies typically are of order 100 Myr, it is immediately clear that age spread effects will be very important -- as first pointed out by Meurer (1995) -- and can significantly distort the shape of the luminosity function with respect to that of the mass function (see FvA 1999). 

Our models also allow to establish theoretical calibrations of arbitrary colors in terms of metallicity [Fe/H], study how they change as a function of SC age and compare to empirical calibrations for Milky Way or M31 GCs (cf. Couture \etal 1990, Barmby \etal 2000). At old ages $\sim 12$ Gyr, our theoretical calibrations agree very well with the empirical ones for the metallicity range of Milky Way and M31 GCs, but significantly steepen at [Fe/H]$>-0.5$. This implies that metallicities derived from linear extrapolations of empirical relations beyond [Fe/H]$=-0.5$ are severely overestimated as comfirmed by spectroscopic observations from Kissler-Patig \etal (1998). The theoretical calibrations appreciably change for ages $\leq 10$ Gyr (cf. Schulz \etal 2002).

\section{A Grid of Star Cluster Models}
Broad and intermediate band colors in any filter system and even spectral features like ${\rm H_{\beta}}$ are to some degree degenerate in terms of age and metallicity. In the presence of dust which plays a role in all environments in which YSCs are observed to form, an additional dimension is added to this degeneracy extending it to an age -- metallicity -- extinction degeneracy. It is worst, as is well known, for optical colors. Optical -- NIR colors already better split up in metallicity, and UV -- optical colors give a better handle on extinction and ages. In any case, multi-wavelength analyses are required and very useful -- as I will show -- to individually determine metallicities, ages and extinction values  for the sometimes rich star clusters populations forming in starbursting and merging galaxies as well as for intermediate-age and old cluster systems. As SC systems age, dust becomes less and less of an issue, in general. The ASTROVIRTEL project {\sl Evolution and Environmental Dependence of Star Cluster Luminosity Functions} (PI: R. de Grijs, CoIs: UFvA, G. Gilmore) provides multi-wavelength photometric data in a homogeneous way from HST, VLT and other archives. I.e., it provides {\bf S}pectral {\bf E}nergy {\bf D}istributions ({\bf SED}s) for star clusters from {\sl UV} through NIR. 
As part of his PhD thesis, P. Anders has developed an algorithm that compares star cluster SEDs with an extensive grid of evolutionary synthesis model SEDs as described above. Dust extinction is added to the model spectra for various ${\rm E_{B-V}}$ using Calzetti \etal's (2000) starburst extinction law. The model grid covers
\begin{itemize}  
\item[-] 5 metallicities in the range ${\rm -1.7 \leq [Fe/H] \leq +0.4}$
\item[-] 1170 ages from 4 Myr all through 14 Gyr and
\item[-] 20 extinction values ${\rm 0 \leq E_{B-V} \leq 1}$
\end{itemize}
resulting in a total of 117.000 SEDs.

\begin{figure}[!h]
\begin{center}
\includegraphics[angle=-90,width=8.cm]{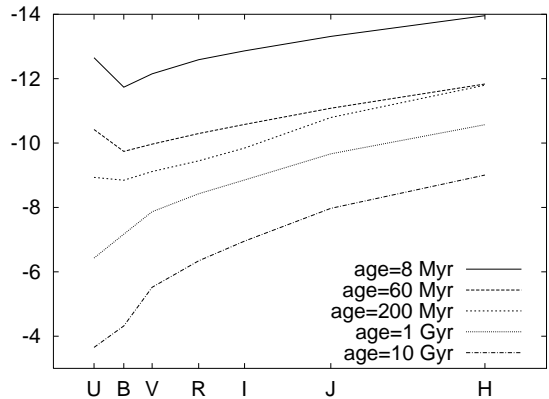}
\end{center}
\caption{Model SEDs for SCs of 5 different ages, solar metallicity and no extinction.}
\end{figure}

Fig. 2 shows model SEDs for [Fe/H]$=0$, ${\rm E_{B-V}=0}$ and 5 different ages as an example. Note the strong change in $ U-$ band magnitudes relative to longer wavelength ones. Increasing or lowering model cluster masses shifts the SED to brighter or fainter magnitudes without affecting its shape.

\section{Analysing Star Cluster Systems}
The SC analysis tool developed by P. Anders compares an observed SC SED with all ${\rm n=117.000}$ SEDs from the model grid and assigns a probability to each model SED by means of a maximum likelihood estimator ${\rm p(n) \sim exp(-\chi^2)}$ with ${\rm \chi^2=\sum_{\lambda}(m_{\lambda}^{obs}-m_{\lambda}^{model})^2/\sigma_{obs}^2}$. Then the ${\rm p(n)}$ are normalised to ${\rm \sum_n p(n)=1}$. The model SED with the highest probability p gives the best agreement with the observed cluster in terms of metallicity, age and ${\rm E_{B-V}}$. Models are summed up in order of decreasing probability until ${\rm \sum p=0.68}$ is reached to yield the $\pm 1 \sigma$ uncertainties in all three parameters (see Anders \etal 2004a for details). The mass of a SC is derived from the amount of vertical shift of the SED. 

When applied to accurate photometry in reasonably chosen passband combinations, this SED analysis tool allows to determine ages, metallicities, extinction value and masses including their $1 \sigma$ uncertainties for arbitrary large SC systems and, hence, to assess e.g. the mass function of a SC system. 

To assess the precision and limitations of our SED analysis tool, extensive tests were carried out on artificial SCs of 5 different ages of 8, 60, and 200 Myr, 1 and 10 Gyr. Studying the effects of the photometric accuracy on the precision to which artificial SC properties can be recovered, it turns out that for young clusters the input properties are very well recovered from multi-band photometry with accuracies $\leq 0.1 ~\dots~ 0.2$ mag. For the oldest clusters, the recovered parameter values age metallicity and extinction start to differ from the input values already for photometric accuracies $\gta 0.1$ mag. 

\begin{figure}[!h]
\begin{center}
\includegraphics[angle=-90,width=12.cm]{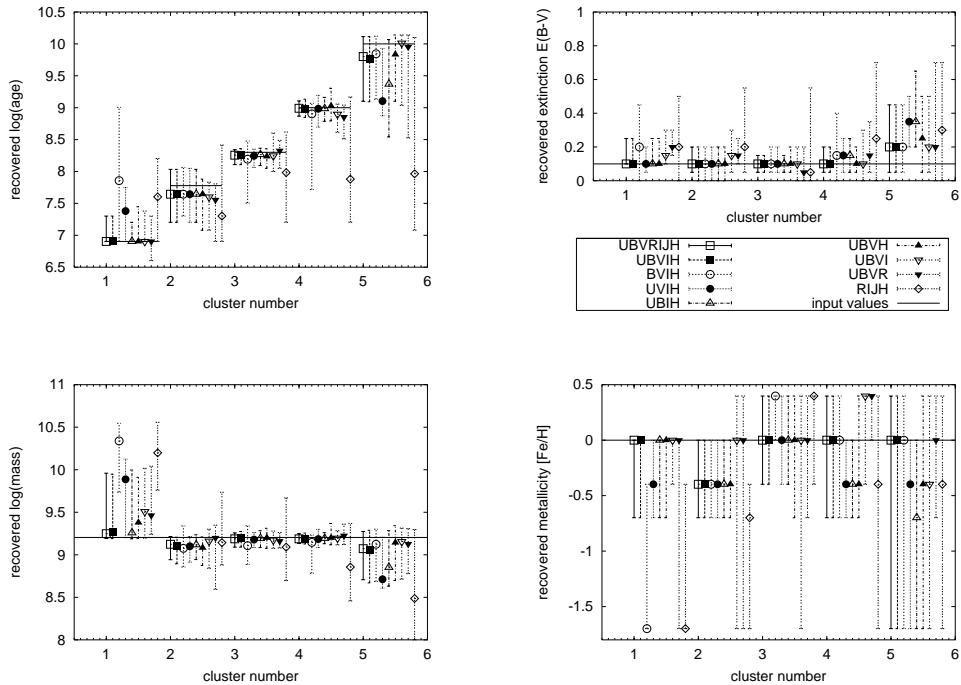}
\end{center}
\caption{Dispersion of properties recovered by our SED analysis tool for artificial SCs of 5 different ages 8, 60 and 200 Myr, 1 and 10 Gyr, [Fe/H]=0, ${\rm E_{B-V}=0.1}$, mass ${\rm 1.6 \cdot 10^9~M_{\odot}}$, and an assumed photometric accuracy of 0.2 mag for different passband combinations.}
\end{figure}

Next, we investigate the precision to which artificial cluster properties are recovered for an assumed photometric accuracy of 0.2 mag if only a limited number of passbands were observed and we investigate which passband combination provides the maximum amount of information. Fig. 3 clearly shows that in order to recover ages, metallicities and extinctions for YSCs the $U-$band is crucial, whereas one NIR-band $J,~H,$ or $K$ is important to determine cluster metallicities. A long wavelength basis from $U-$ through NIR is essential. Additional photometry in a second or third NIR band does not provide much additional information or precision. 

We also investigated the impact of {\sl a priori} assumptions, like e.g. asuming solar metallicity for YSC as often found in the literature, on the results of our SED analyses. Much like the simple comparison between observed colors and SSP models, our SED analysis tool does find solutions, i.e. ages and extinction values, in this case. However, these differ substantially and systematically from the true values if the SCs in fact have slightly lower metallicity. Already for SC metallicities around ${\rm \sim \onehalf~Z_{\odot}}$ the assumption of solar metallicity leads to ages underestimated by factors $2-8$ for SC ages $\lta 100$ Myr (see Anders \etal 2004a for a more extensive discussion). 

We conclude from artificial SC experiments that from reasonably accurate ($\lta 0.2$ mag) photometric data over a long enough wavelength basis with a minimum of 4 filters ($U, B, V$ or $I, H$ or $K$) our SED analysis tool allows to independently determine ages, metallicities, extinction values and, hence, masses of YSCs with high precision: ages to $\lta 20\%$, metallicities to $\lta 0.2$ dex, ${\rm E_{B-V}}$ to $\lta 0.05$ mag, and masses to $\lta 25 \%$ for SCs with ages $\lta 200$ Myr. We recall that this is possible with HST WFPC2 $+$ NICMOS or HST WFPC2 in combination with VLT ISAAC for rich cluster systems in only 4 reasonably deep exposures. 

Interestingly, Cardiel {\it et al.} (2003) show that for typical photometric accuracies, broad band photometry with useful passband combinations ({\sl VIK}) is as powerful in disentangling ages and metallicities of old stellar populations as is spectroscopy with typical S/N per ${\rm \AA}$ ratios.

\subsection{A First Application: NGC 1569}
As a 1$^{\rm st}$ example, our SED analysis tool was applied to the SC system in the nearby (D$\sim 2.2$ Mpc) post-starburst dwarf galaxy NGC 1569 with initially two well-known Super Star Clusters (now 3 as one turned out to be composed of 2 different stellar populations). The data provided by our ASTROVIRTEL project encompass reduced and homogeneously calibrated HST F336W, F380W, F439W, F555W, F814W, F110W, F160W passband observations. Anders \etal (2004b) identified a rich system of $\gta 160$ SCs and determined their individual ages, metallicities, ${\rm E_{B-V}}$, and masses. The bulk of the SCs have ages $\leq 25$ Myr and significantly subsolar metallicities, only 15 \% are older. Extinction in NGC 1569 is $\leq 0.1$ for 73\% of the SCs, only 10 of the SCs ($=7\%$ and primarily the very youngest ones) have ${\rm E_{B-V} \geq 0.5}$. Except for the 3 Super Star Clusters, only 3 other clusters reach masses ${\rm >10^5~M_{\odot}}$ as typical for Galactic GCs. SC masses in NGC 1569 typically range from $\sim 10^3$ to several ${\rm 10^4~M_{\odot}}$, instead. Subdivision of the cluster mass function into age bins indicates that either the early burst stages formed a larger number of massive clusters than the later ones or that the less massive clusters that did form early in the burst were already destroyed within the 25 Myr since then.

The most important result of this study is that the starburst in the isolated dwarf galaxy NGC 1569 apparantly did {\bf not} form many GCs -- at variance with what happens in starbursts accompanying mergers of massive gas-rich galaxies. The strong starburst occuring $\sim 600 - 800$ Myr ago during the merger of 2 massive gas-rich spirals now called NGC 7252, in contrast, did form a rich population of SCs with masses typical of GCs and apparently strongly enough bound to have survived the violent relaxation processes in this merger. This raises the interesting question if it is the shallower potential, lower velocity dispersion, smaller gas reservoir, shorter dynamical timescale, etc. of a dwarf galaxy on the one hand or the absence of an external trigger and the high pressure typical for massive galaxy mergers on the other hand that prevented the starburst in NGC 1569 from forming a rich population of young GC-like objects. 

\section{Conclusions and Perspectives}
We have shown that with 3 ingredients
\begin{itemize}
\item[-] evolutionary synthesis models using {\bf complete} sets of stellar evolutionary tracks for different metallicities and consistently accounting for gaseous emission contributions in early stages to produce an extensive grid of SEDs from UV through NIR
\item[-] broad band photometry in 4 bands ({\sl U, B, V} or {\sl I, H} or {\sl K}) with photometric accuracy $\leq 0.2$ mag
\item[-] an analysis tool for SEDs as presented here 
\end{itemize}
it is possible to independently determine ages, metallicities, extinction values and masses including their respective $\pm 1 \sigma$ uncertainties for large numbers of individual SCs in external galaxies. I have also shown that in particular for young star clusters considerable precision is reached for photometric accuracies $\lta 0.2$ mag : $\lta 20$\% in ages, $\lta 0.2$ dex in metallicity, $\lta 0.05$ mag in ${\rm E_{B-V}}$, and $\lta 25$ \% in masses. Star cluster systems older than $\sim 200$ Myr require better photometric accuracy than that or else have somewhat larger uncertainties in their derived parameters.

We cautioned that empirical GC color -- metallicity calibrations are only valid for ages $> 10$ Gyr and [Fe/H]$<-0.5$. Theoretical calibrations can be obtained from evolutionary synthesis models for arbitrary cluster ages. They indicate that metallicities derived from linear extrapolations of empirical relations significantly overestimate true cluster metallicities. 

The strong age dependence of M/${\rm L_V-}$ values at early stages leads to considerable distortions of cluster luminosity functions with respect to mass functions.

{\sl A priori} assumptions like e.g. solar metallicity for YSCs -- even if not too far from their true metallicity -- lead to significant and systematic errors in cluster ages and masses. 

An ESO ASTROVIRTEL project provided us with a wealth of multi-wave\-length photometric data for star cluster systems in starburst and interacting galaxies, ongoing mergers and merger remnants, dynamically young and fiducially old E/S0s. As exemplified for NGC 1569 we will use our SED analysis tool to determine individual cluster ages, [Fe/H], ${\rm E_{B-V}}$, and masses, study and compare mass functions of young, intermediate-age and old star cluster systems to understand cluster formation and destruction processes, understand ages and metallicities of the blue and red peak GC subpopulations in E\S0 galaxies. Because star clusters can be studied one-by-one, star cluster systems are much better suited than integrated light for studies of the violent (star) formation histories of their parent galaxies in order to assess in detail cosmological galaxy formation scenarios, redshifts and timescales.

\acknowledgements
Deep thanks to my collaborators on this project Peter Anders and Richard de Grijs.  
We gratefully acknowledge support by ASTROVIRTEL, funded by the European
Commission under HPRI-CT-1999-00081, and partial financial support from the DFG under Fr 916/11-1. It's a pleasure to thank the organisers for a very stimulating conference and to gratefully acknowledge their travel support.

\end{document}